\documentclass[12pt,notitlepage,aps,preprintnumbers,superscriptaddress,nofootinbib,aps,prd,floatfix]{revtex4-2}
\usepackage{float}
\usepackage{physics,slashed}
\usepackage{amsfonts,amsmath,amssymb}
\usepackage{mathrsfs}
\usepackage{bm}

\usepackage{epsfig}
\usepackage{graphicx}   
\usepackage{subfigure}  
\usepackage{verbatim}   
\usepackage{color}      
\usepackage{hyperref}   
\usepackage{url}
\definecolor{nicered}{rgb}{0.5,0.,0.}
\definecolor{nicegreen}{rgb}{0.,0.5,0.}
\definecolor{niceblue}{rgb}{0.,0.,0.5}
\hypersetup{colorlinks,citecolor=nicegreen,linkcolor=nicered,urlcolor=niceblue}
\usepackage{array}
\usepackage{dcolumn}
\usepackage{multirow}
\usepackage{cprotect}
\usepackage{framed}
\usepackage{setspace}

\newcommand{\TeV}{\textrm{TeV}}
\newcommand{\GeV}{\textrm{GeV}}

\newcommand{\bea}{\begin{equation}\begin{aligned}}
\newcommand{\eea}{\end{aligned}\end{equation}}

\begin{document}
\title{Electroweak fragmentation at high energies:\\ A Snowmass White Paper}
\preprint{PITT-PACC-2204, Snowmass 2021 TF07\&EF04}
\author{Tao Han}\email{than@pitt.edu}
\author{Yang Ma}\email{mayangluon@pitt.edu}
\author{Keping Xie}\email{xiekeping@pitt.edu}
\affiliation{Pittsburgh Particle Physics, Astrophysics,and Cosmology Center, 
Department of Physics and Astronomy, University of Pittsburgh, Pittsburgh, PA 15206, USA\looseness=-1}

\date{\today}
\begin{abstract}
In particle collisions at energies higher than the mass scale involved, the collinear splitting becomes the dominant phenomena. Suitable descriptions of the  physics in this regime include the parton distribution functions, the initial state radiations, the final state radiations and the  fragmentation functions. It is of fundamental importance to formulate those functions with consistent theoretical treatments, and to provide the adequate formalism for applications, as  motivated by the recent discussions of partonic scatterings at the multi-TeV energy regime. In this report, we briefly present the resummation of the final-state logarithms as fragmentation functions for the electroweak processes. As an explicit example, we study an electroweak gauge boson splitting in the process  $pp\to WZj$ and demonstrate the important effects from the fragmentation.

\end{abstract}

\singlespacing
\maketitle

\section{Introduction}

In particle collisions at energies much larger than the mass scale involved, the collinear splitting becomes the dominant phenomena. The physics in this regime is described by parton distribution functions (PDFs), initial state radiations (ISRs), final state radiations (FSRs), and fragmentation functions (FFs).  
%
At an energy scale well above the electroweak (EW) scale $(\mu_{\rm EW}\sim M_Z)$, the Standard Model (SM) particles, including $W,Z,H,t$, can be treated essentially massless and 
the electroweak gauge symmetry 
$SU(2)_L \times U(1)_Y$ is largely restored. 
Under such a circumstance, the collinear splitting phenomena
dominate due to the large logarithmic enhancement \cite{Chen:2016wkt}, in the form $\alpha_I\log(Q^2/M^2)$, where $\alpha_I$ is the involved coupling, $Q^2$ is the factorization scale, identified as the typical momentum transfer of the physical process. As such, the convergence of the fixed-order calculation becomes inadequate and as higher-order corrections as powers  $\alpha_I^m\log^n(Q^2/M^2)$ should be resummed in order to improve the  theoretical prediction.

Generally speaking, the large logarithms in intial-state raditions (ISRs) should be resummed as parton distribution functions (PDFs), and the ones in final-state radiations (FSRs) are to be treated similarly as fragmentation functions (FFs), leading to the $Q^2$-dependent functions governed by the well-known DGLAP equations  \cite{Altarelli:1977zs,Gribov:1972ri,Lipatov:1974qm,Dokshitzer:1977sg}. 
The full evolution of the SM PDFs (including both EW and QCD ones) is completed recently \cite{Bauer:2017isx,Bauer:2018arx,Han:2020uid,Han:2021kes,Buarque:2021dji} and their direct application is motivated by the discussions of future 100 TeV proton-proton colliders \cite{Tang:2015qga,FCC:2018vvp} and a multi-TeV muon collider \cite{Schulte:2020xvf}. Specific  implications at a high-energy lepton collider are investigated with a few typical SM processes~\cite{Han:2020uid,Han:2021kes,Buarque:2021dji}. 
In this report, we extend our work to FFs, to the leading log accuracy for the electroweak processes. As an explicit example, we study an electroweak gauge boson ($V$) splitting in the process $pp\to WZj$. Some related works have already appeared in Refs.~\cite{Bauer:2018xag,Bauer:2020jay}.

\section{The theoretical framework}

Consider a generic $2\to 3$ scattering process at high energies. The cross section may be factorized as a convolution between a partonic $2\to 2$ cross section and a universal $1\to 2$ splitting function, formally written as
\begin{equation}
\sigma_{2\to3}= \hat{\sigma}_{2\to2} \otimes 
P_{1\to 2},
\end{equation}
where $\hat{\sigma}_{2\to2}$ denotes the partonic cross section, and $P_{1\to2}$ stands for the corresponding splitting function. 
The collinear splitting in ISRs can be resummed into the parton distribution functions $f_{i/{\rm beam}}$, defined as the probability of finding the parton $i$ inside of the beam particle. 
Similarly, we can define the fragmentation (or decay) functions (FFs) $d^{f}_{i}$, which ascribe the FSR splittings $i\to f$. 
Similar to the PDF evolution, FFs also run with energy scale $Q$, in terms of DGLAP equations~\cite{Altarelli:1977zs,Gribov:1972ri,Lipatov:1974qm,Dokshitzer:1977sg},
\begin{equation}
\label{eq:DGLAP}
\pdv{}{\log Q^2} d^{f}_{i}(x,Q^2)=
\sum_{I}\frac{\alpha_I}{2\pi}
\sum_{j}\int_{x}^1\frac{\dd z}{z} d^{f}_{j}(z,Q^2)P^{I}_{ji}(x/z),
\end{equation}
where the index $I$ runs over all the SM gauge and Yukawa interactions.
The splitting kernel $P_{ji}$ corresponds to $i$ splitting into $j$.
Similarly to Ref.~\cite{Bauer:2018xag}, we take the $Q_0=\mu_{\rm EW}$ as the starting scale and run upwardly.
The solution, $d^{f}_{i}(x,Q^2)$, should be interpreted as probability of finding a definite particle $f$ at the scale $Q_0=\mu_{\rm EW}$ from a mother particle $i$ produced at the scale $Q>Q_0$. 

\begin{figure}
\centering
\includegraphics[width=0.49\textwidth]{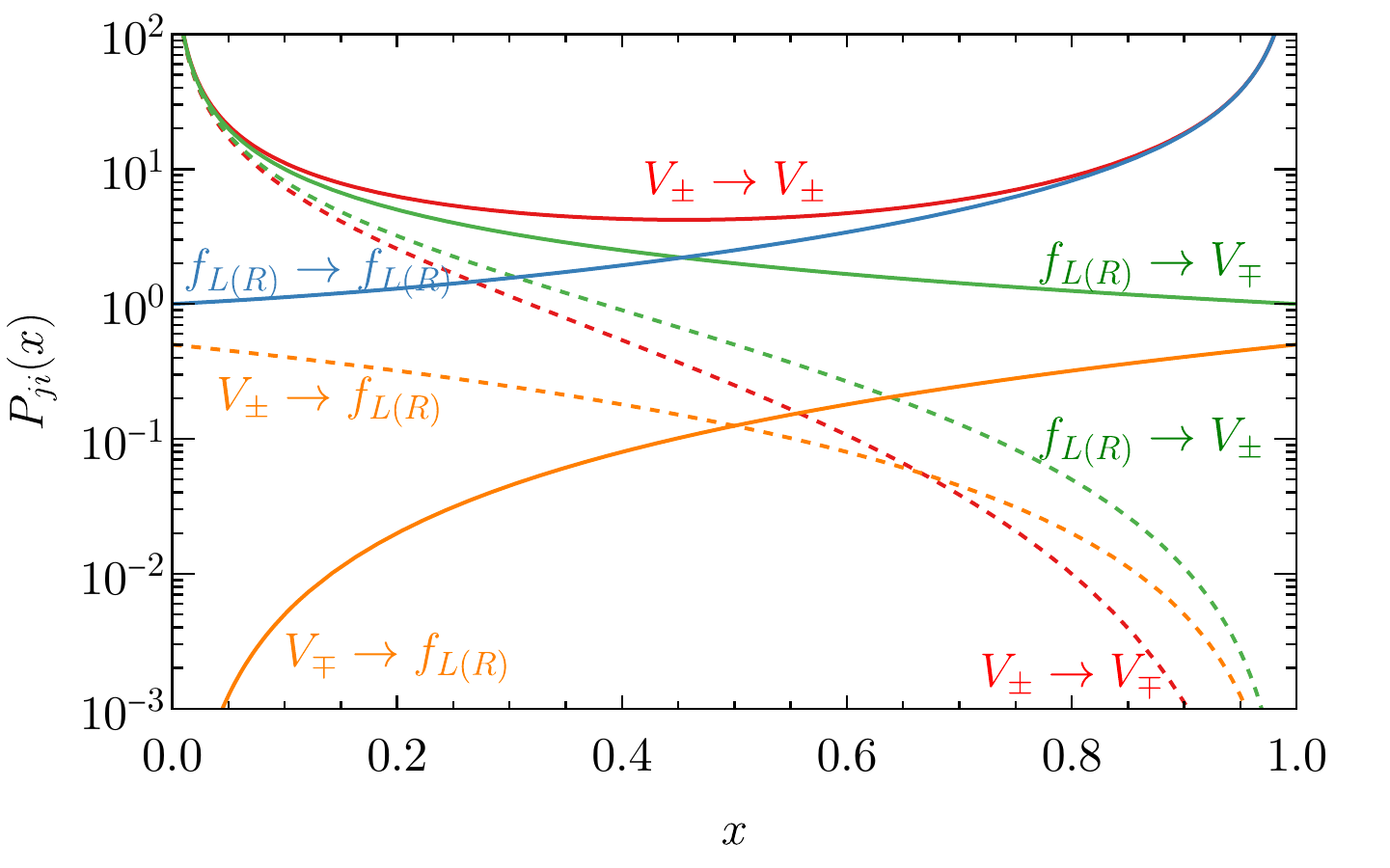}
\includegraphics[width=0.49\textwidth]{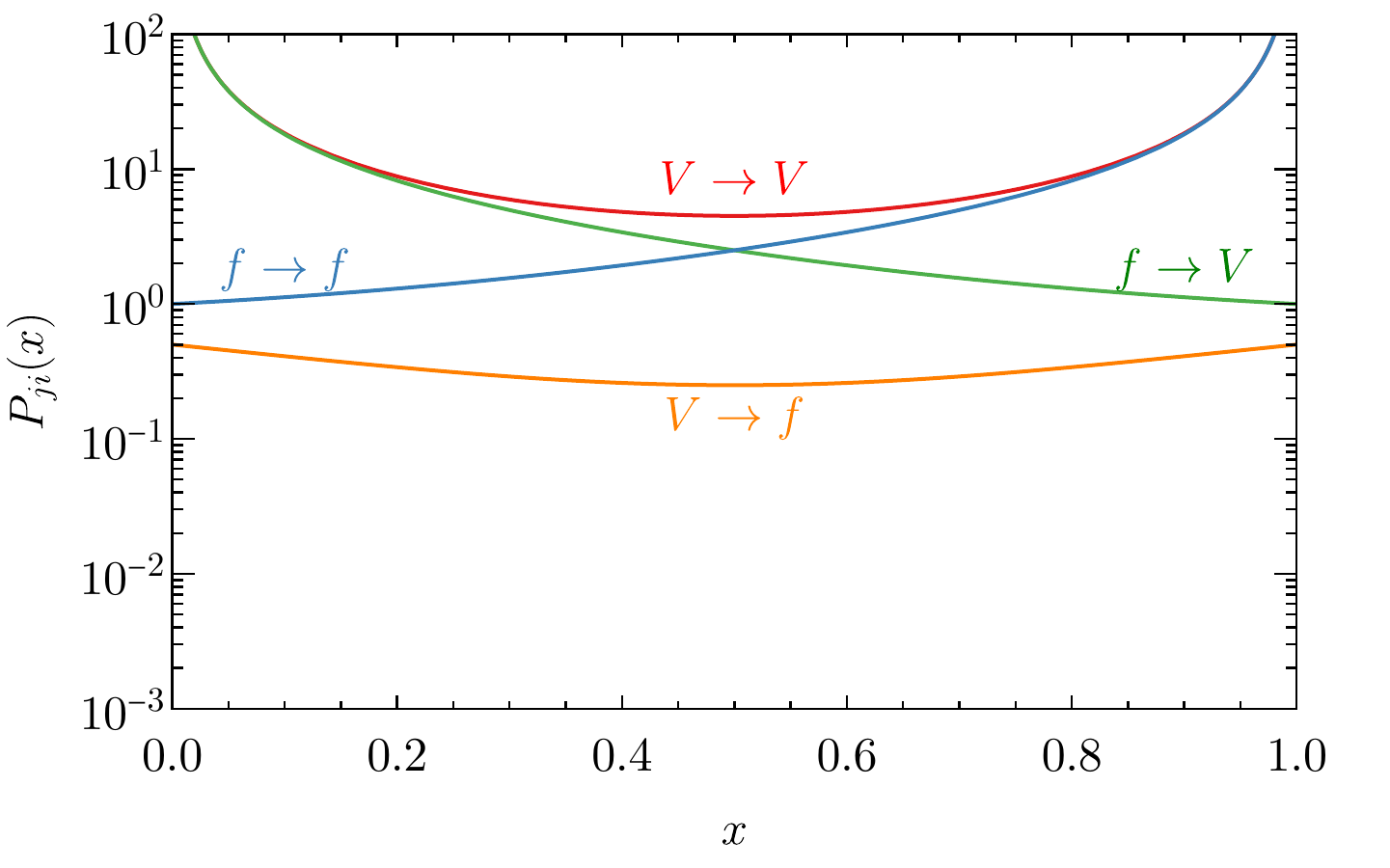}
\caption{The relevant polarized (left) and unpolarized (right) splitting functions $P_{ji}$. 
}
\label{fig:split}
\end{figure}

We first present some typical splitting functions $P_{ji}$ for the processes $i\to j$ in Fig.~\ref{fig:split} with polarized massless fermions $f_{L,R}$ and vector bosons $V_\pm$ (left) and unpolarized (right) states.\footnote{The notation $f_{L(R)}\to V_{\pm}$ indicates $f_L\to V_+$ and $f_{R}\to V_{-}$, which share the same form due to the charge-parity transformation.} The notable feature is the infrared behavior of a gauge boson at $x \to 0$. 
We next evaluate the fragmentation functions $d^f_{i}(x,Q^2)$ for the processes $i\to f$ as in Eq.~(\ref{eq:DGLAP}). 
The initial conditions for vector bosons ($V=W^\pm,Z$) can be taken as
\begin{equation}
d^{V}_{V_{+}}(x,Q_0^2)=d^{V}_{V_{-}}(x,Q_0^2)=\delta(1-x),~ d^{V}_{i\neq V}(x,Q_0^2)=0,
\end{equation}
Similarly, for fermions, we have
\begin{equation}
    d^{f}_{f_L}(x,Q_0^2)=d^{f}_{f_R}(x,Q_0^2)=\delta(1-x), ~d^{f}_{i\neq f}(x,Q_0^2)=0.
\end{equation}
We assume that the vector boson polarizations ($\pm$) and the fermion helicities ($L,R$) are unobservable. We only have left-hand neutrinos, with the corresponding input as
\begin{equation}
    d^{\nu}_{\nu_L}(x,Q_0^2)=\delta(1-x), ~d^{\nu}_{i\neq\nu}(x,Q_0^2)=0.
\end{equation}

We take $i\to W^+$ as an explicit example to illustrate the main features of splittings, with implications shown explicitly later in next section.
We can solve the DGLAP equation iteratively, as developed in Refs.~\cite{Han:2020uid,Han:2021kes,Buarque:2021dji}. 
For simplicity, we take fixed couplings
\begin{equation}
\alpha_1=0.01, ~\alpha_2=0.0348, ~\alpha_3=0.118,
\end{equation}
determined through the PDG world average values~\cite{ParticleDataGroup:2020ssz}:
\begin{equation}
\alpha_s(M_Z^2)=0.118,~\alpha_e(M_Z^2)=1/128.8~, M_{W(Z)}=80.379~(91.1876)~\GeV.
\end{equation}
The fixed couplings are justified by their small variations within a multi-TeV range, $\mu_{\rm EW}<Q\lesssim$ a few TeV. Future improvements can include running couplings as well, as demonstrated in Ref.~\cite{Han:2021kes}.

\begin{figure}
\centering
\includegraphics[width=0.8\textwidth]{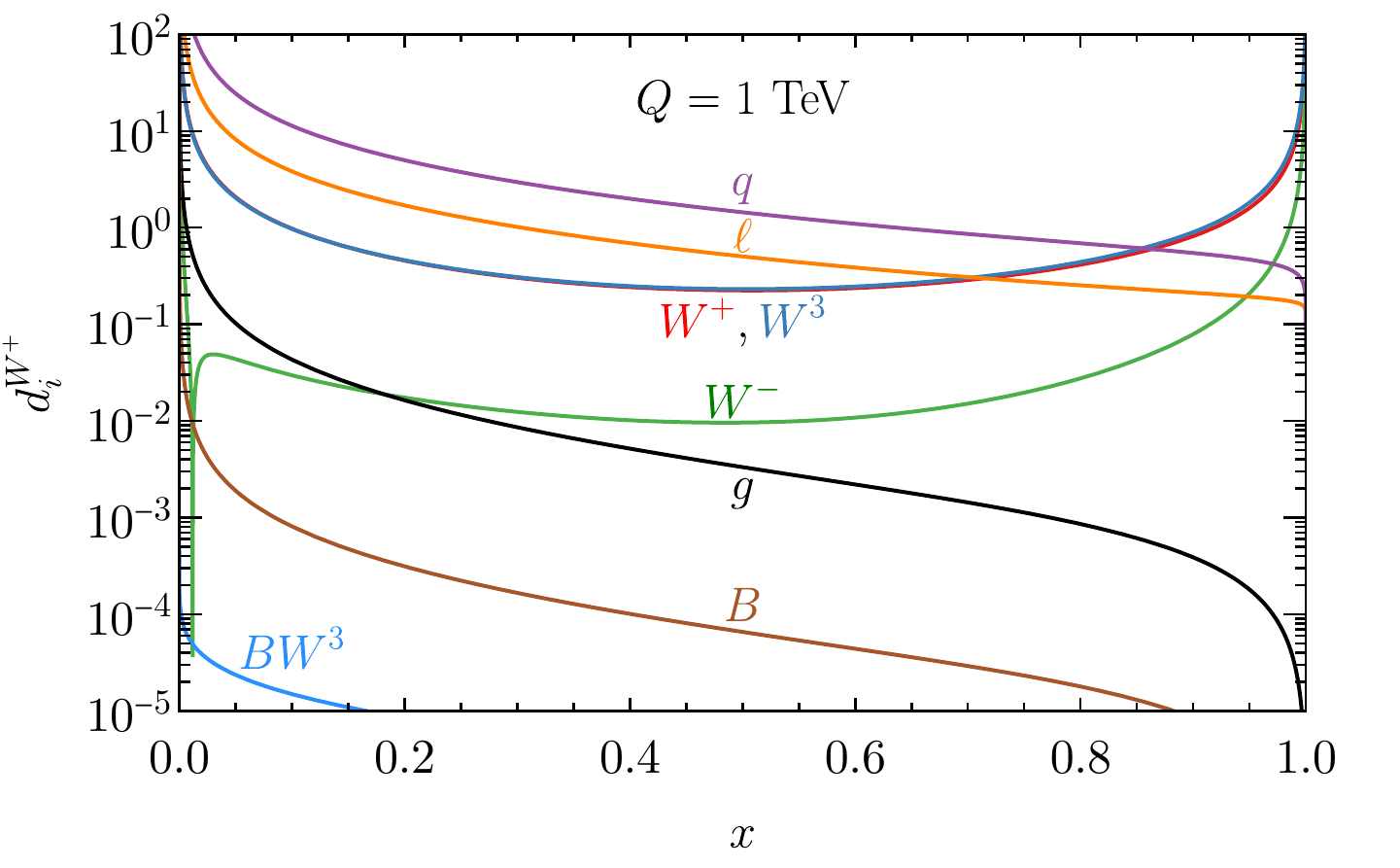}
\caption{Fragmentation functions of $d^{W^+}_{i}(x,Q^2)$ in the splittings $i\to W^+$, with the initial state $i=q,g,\ell,W^{\pm,3},B$ and the mixed state $BW^3$.}
\label{fig:dW}
\end{figure}

In Fig.~\ref{fig:dW}, we display the numerical results of FFs $d^{W^+}_{i}$ from various mother particles $i$ in the unbroken gauge basis, produced at an energy scale $Q=1~\TeV$. Several remarks are in order. \\
%
$\bullet$ 
We see the FFs from the mother particle $W^+$ and $W^{3}$ are peaked around both $x\to1$ and $x\to0$, as a result of the infrared behavior from the vector boson splittings $W^+\to W^+W^3$ and $W^3\to W^+W^-$.\\
$\bullet$
In dealing with $d^{W^+}_{\ell}$ and $d^{W^+}_{q}$ from fermion splittings, we have summed over the flavors for simplicity  
\begin{equation}\label{eq:sum}
q=\sum_{i=1}^{6}(q_i+\bar{q}_i), \quad 
\ell=\sum_{i=1}^{6}(\ell_i+\bar{\ell}_i),
\end{equation}
with the lepton component including neutrinos as well.
The size of the neutrino FFs ($\nu\to W^+\ell^-$) is roughly the same as the  charged-lepton one ($\ell^+\to W^+\bar{\nu}$), with a tiny difference resulted from higher-order splittings involving hyper charges. The sum of quark FFs $d^{W^+}_{q}$ is roughly three times of the lepton ones $d^{W^+}_{\ell}$, as a consequence of the additional color factor. \\
$\bullet$ 
The gluon contributions enter as $g\to q\bar{q}$, followed by $q\to W^+ q'$. It is at the sub-leading order $\alpha_3\alpha_2$. \\
$\bullet$
At a sub-leading splitting, we begin to have mother particle of $W^-$, mostly due to the splitting chain $W^{-}\to W^{-}W^3$ followed by $W^3\to W^+W^-$. It is important to note that at an extremely small momentum fraction, $x<\mu_{\rm EW}/Q$, the $W^-$ FF will flip sign, shown as a sharp dip in Fig.~\ref{fig:dW}. It is a result of incomplete cancellation of the infrared divergence, due to the Bloch-Nordsieck theorem violation~\cite{Chen:2016wkt,Ciafaloni:2001mu}. \\
$\bullet$
We also have the hyper-charge gauge boson $B$ from $B\to f\bar{f}$, followed by $f\to W^+f'$. The mixed-state (coherent) component $d^{W^+}_{BW^3}$ is highly suppressed, which only shows up in the lower-left corner in Fig.~\ref{fig:dW} with two magnitudes smaller than $d^{W^+}_{B}$. \\
$\bullet$
Numerically, we note that our results are highly stable in the shower expansion. We only iterate a few steps (up to $3\sim5$), which are sufficient for the convergence of FFs in the EW sector. 
We want to emphasize that the QCD sector $d^{W^+}_{q}$ and $d^{W^+}_g$ may receive sizable corrections from higher orders of the QCD interaction, which are left for a future exploration.

The representation in the physical basis, more appropriate in realistic calculations in the broken phase, can be easily obtained through the rotation 
\begin{equation}
\begin{pmatrix}
d_{\gamma}\\ 
d_{Z}\\ 
d_{\gamma Z}
\end{pmatrix}=\begin{pmatrix}
c_W^2 & s_W^2  & c_W s_W\\
s_W^2 & c_W^2  & -c_W s_W\\
-2c_W s_W & 2c_W s_W & c_W^2-s_W^2 
\end{pmatrix}
\begin{pmatrix}
d_{B} \\ 
d_{W^3}\\ 
d_{BW^3}
\end{pmatrix},
\end{equation}
where $s_W=\sin\theta_W,\ c_W=\cos\theta_W$,  with respective to the weak mixing angle $\theta_W$. 

\section{An explicit fragmentation example}
\begin{figure}
\centering
\includegraphics[width=0.3\textwidth]{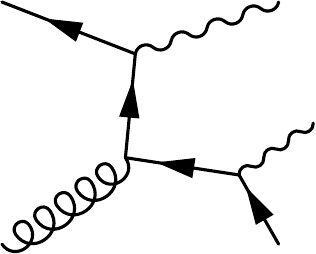}
\includegraphics[width=0.3\textwidth]{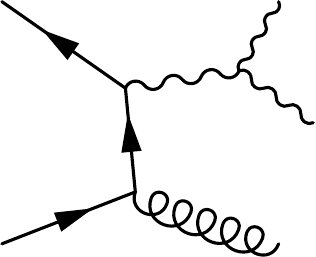}
\caption{The representative Feynman diagrams for high-$p_T$ $pp\to VVj$ events from the splittings $q\to q V$ and $V\to VV$.}
\label{fig:feyn}
\end{figure}
At high energies, events with gauge bosons are dominated by the final state splittings.
We take the high-$p_T$ events in the $pp\to VVj\ (V=W^\pm,Z)$ production as an example to illustrate the collinear splitting in the final state radiations (FSRs). 
The representative Feynman diagrams, with $q\to qV$ and $V\to VV$ splittings, are shown in Fig.~\ref{fig:feyn}.
The dominant contribution to the cross section is originated from a final-state collinear splitting, 
formally written as
\begin{equation}\label{eq:FSR}
\sigma_{\rm FSR}\supset\hat{\sigma}(pp\to Vq)\otimes P(q\to qV)
+\hat{\sigma}(pp\to Vj)\otimes P(V\to VV).
\end{equation}
In the fragmentation approach, the splitting functions $P(q\to qV)$ and $P(V\to VV)$ need to be resummed as the corresponding FFs, as discussed in the previous section. As an alternative approach, the resummation can be done with Sudakov form factor, simulated with parton shower, as widely adopted in the simulation packages such as in PYTHIA~\cite{Christiansen:2014kba} or HERWIG~\cite{Masouminia:2021kne}.

\begin{figure}
    \centering
    \includegraphics[width=0.49\textwidth]{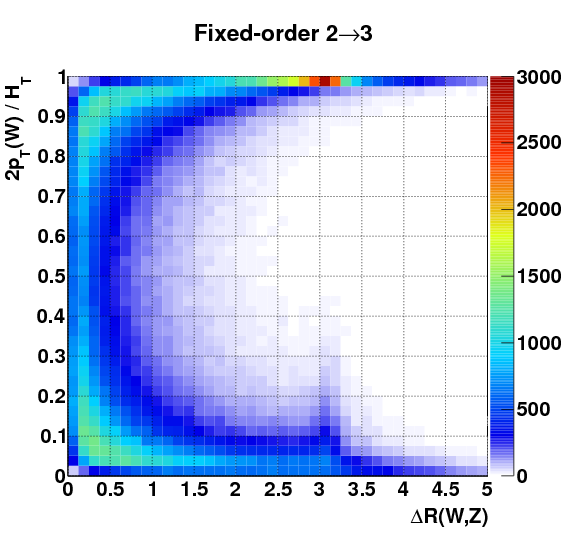}    
    \includegraphics[width=0.49\textwidth]{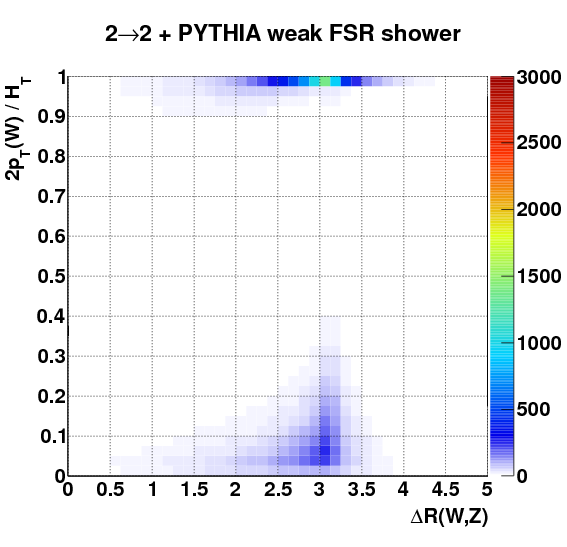}
    \includegraphics[width=0.49\textwidth]{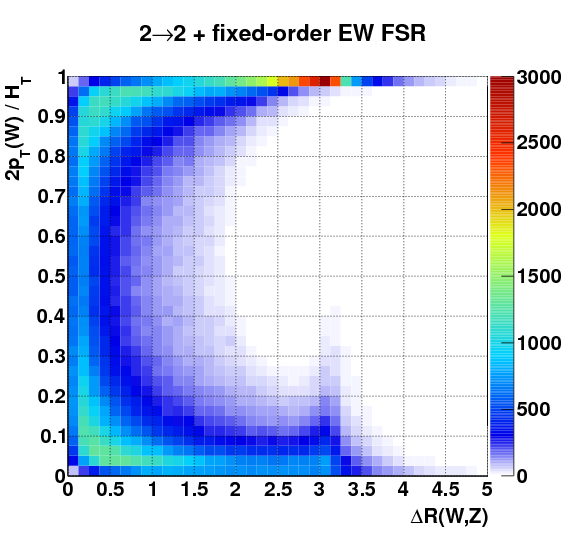}    
    \includegraphics[width=0.49\textwidth]{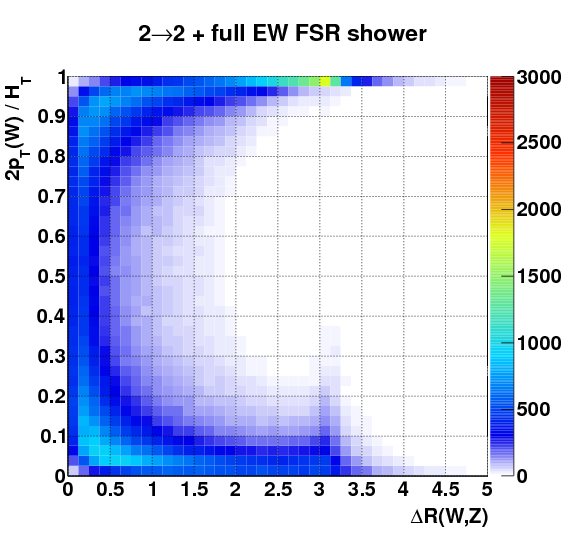}
    \caption{The relative event population for the $WZj$ production at a 100 TeV $pp$ collider in the plane of $2p_T^{W}/H_T$ versus $\Delta R_{WZ}$, with $p_T^j>3$ TeV.
    Upper left: fixed order calculation for $2\to3$ $WZj$ production, generated with MadGraph~\cite{Alwall:2014hca};
    upper right: PYTHIA weak shower, which include $q\to qV$~\cite{Christiansen:2014kba}; 
    Lower left: the fixed-order EW FSR, including $V\to VV$ splitting as well. Lower right: the full FSR shower. Plots adapted from Ref.~\cite{Chen:2016wkt}.
    }
    \label{fig:EWshower}
\end{figure}

If we focus more specifically on the $pp\to WZj$ production, a large fraction of events are originated from $pp\to Wj$, with subsequent splittings involving $W\to WZ$ and $q\to qZ$. In Fig.~\ref{fig:EWshower}, we show the comparison of the event population of $WZj~(p_T^j>3~\TeV)$ production at a future 100 TeV proton-proton collider, such as SppC~\cite{Tang:2015qga} or FCC~\cite{FCC:2018vvp}, in the plane of $2p_T^W/H_T$ versus the separation distance between $W$ and $Z$ bosons, $\Delta R_{WZ}$, where $H_T$ is defined as scalar sum of all final-state $p_T$s. 
We start with the fixed-order $2\to 3$ calculation done with MadGraph~\cite{Alwall:2014hca}, which should be a good perturbative treatment, as shown in Fig.~\ref{fig:EWshower}(a) (upper-left). 
If we start with the fixed-order perturbative $2\to 2$ calculation for $pp\to Wj$ instead, and then perform the electroweak showering
by  PYTHIA, where only the quark splitting $q\to qV$ is included, we show the results in Fig.~\ref{fig:EWshower}(b) (upper-right).  
We see the PYTHIA captures the feature of event population around the $(\Delta R_{WZ},2p_T^W/H_T)\sim(\pi,0)$ and $(\pi,1)$, 
where $W,Z$ travels in back-to-back directions.
However, in the $\Delta R_{WZ}\to 0$ limit mainly as a cause of $W\to WZ$ splitting, PYTHIA gives a bad description, because of the missing non-Abelian $V\to VV$ splitting in its shower procedure. 
By including this splitting back, we see the fixed-order EW FSR in Fig.~\ref{fig:EWshower}(c) (lower-left)
reproduces the MadGraph fixed-order calculations very well. Going beyond, the full EW FSR shower as in Fig.~\ref{fig:EWshower}(d) (lower-right) includes higher orders of showering as well~\cite{Chen:2016wkt}. 

In general, more work is needed to consistently treat the fragmentation and parton showers, both in theoretical considerations and phenomenological applications. To accomplish the full factorization and to make a more realistic prediction, we need to combine the fragmentation functions with the corresponding partonic cross sections, as indicated in Eq.~(\ref{eq:FSR}). Currently, no public packages are available to easily incorporate the fragmentation functions to obtain the final cross section.
One obstacle we can foresee is the large number of possible intermediate particles, which makes the exhaustion tedious and effort expensive, even though not impossible. 
Instead of the exhausting all intermediate contributions, the more practical approach that we can learn from the parton shower approach \cite{Christiansen:2014kba,Chen:2016wkt,Masouminia:2021kne} is to re-weight the corresponding hard events with the corresponding Sudakov form factor, which equivalently resums the large logarithms in the splitting~\cite{Campbell:2017hsr}. 
To complete the SM framework at high energies, another necessary step is to include the longitudinal gauge bosons and the Higgs boson, specially with large Yukawa coupings for heavy fermions. 
Improved treatments of the longitudinal gauge bosons and their counterparts of the Goldstone bosons need to be properly accommodated in practice \cite{Chen:2016wkt,Cuomo:2019siu}.

\section{Executive summary}
In particle collisions at energies higher than the mass scale involved, the collinear splitting becomes the dominant phenomena. Suitable descriptions of the  physics in this regime include the parton distribution functions (PDFs), the initial state radiations (ISRs), the final state radiations (FSRs) and the fragmentation functions (FFs). It is of fundamental importance to formulate those functions with consistent theoretical treatments, and to provide the adequate formalism for applications. In this report, we focused on fragmentation functions to resum the FSR logarithms up to the leading log (LL) accuracy via the DGLAP equations for the electroweak processes, as shown in Fig.~\ref{fig:dW}. 
As an explicit example, we take the high-$p_T$ process $pp\to WZj$ to demonstrate the important effects of 
the high-energy splittings at a 100 TeV hadron collider. Our formalism is applicable and motivated by the recent discussions for multi-TeV muon colliders. 

Further investigations of the phenomenological applications of EW PDFs and EW FFs at high-energy colliders are under investigation.
It would be also of great importance to combine a comparative study with the resummation through the Sudakov form factor, simulated with parton shower~\cite{Christiansen:2014kba,Chen:2016wkt,Masouminia:2021kne}.

\bibliographystyle{utphys}
\bibliography{ref}
\end{document}